# An SOA Based Design of JUNO DAQ Online Software

Jin Li, *Member, IEEE,* Minhao Gu, Fei Li, *Member, IEEE,* Kejun Zhu, *Senior Member, IEEE*

***Abstract*—The Online Software, manager of the JUNO data acquisition (DAQ) system, is composed of many distributed components working coordinately. It takes the responsibility of configuring, processes management, controlling and information sharing etc. The design of service-oriented architecture (SOA) which represents the modern tendency in the distributed system makes the online software lightweight, loosely coupled, reusable, modular, self-contained and easy to be extended. All the services in the SOA distributed online software system will send messages each to another directly without a traditional broker in the middle, which means that services could operate harmoniously and independently.**

**ZeroMQ is chosen but not the only technical choice as the low-level communication middle-ware because of its high performance and convenient communication model while using Google Protocol Buffers as a marshaling library to unify the pattern of message contents. Considering the general requirement of JUNO, the concept of partition and segment are defined to ensure multiple small-scaled DAQs could run simultaneous and easy to join or leave. All running data except the raw physics events will be transmitted, processed and recorded to the database. High availability (HA) is also taken into account to solve the inevitable single point of failure (SPOF) in the distribution system. This paper will introduce all the core services' functionality and techniques in detail.**

## I. Introduction

THE Jiangmen Underground Neutrino Observatory (JUNO) [1] is a multipurpose neutrino experiment under construction in South of China. It is designed to determine neutrino mass hierarchy, precisely measure oscillation parameters and carry out many other frontier scientific researches by detecting reactor neutrinos from the Yangjiang and Taishan Nuclear Power Plants in Jiangmen, China as shown in Fig.1. JUNO [1] will be the largest liquid scintillator detectors (20 kton) and it will make use of about 20000 20'' photomultiplier tubes (PMT) and 25000 3'' PMTs providing an unprecedented energy resolution.

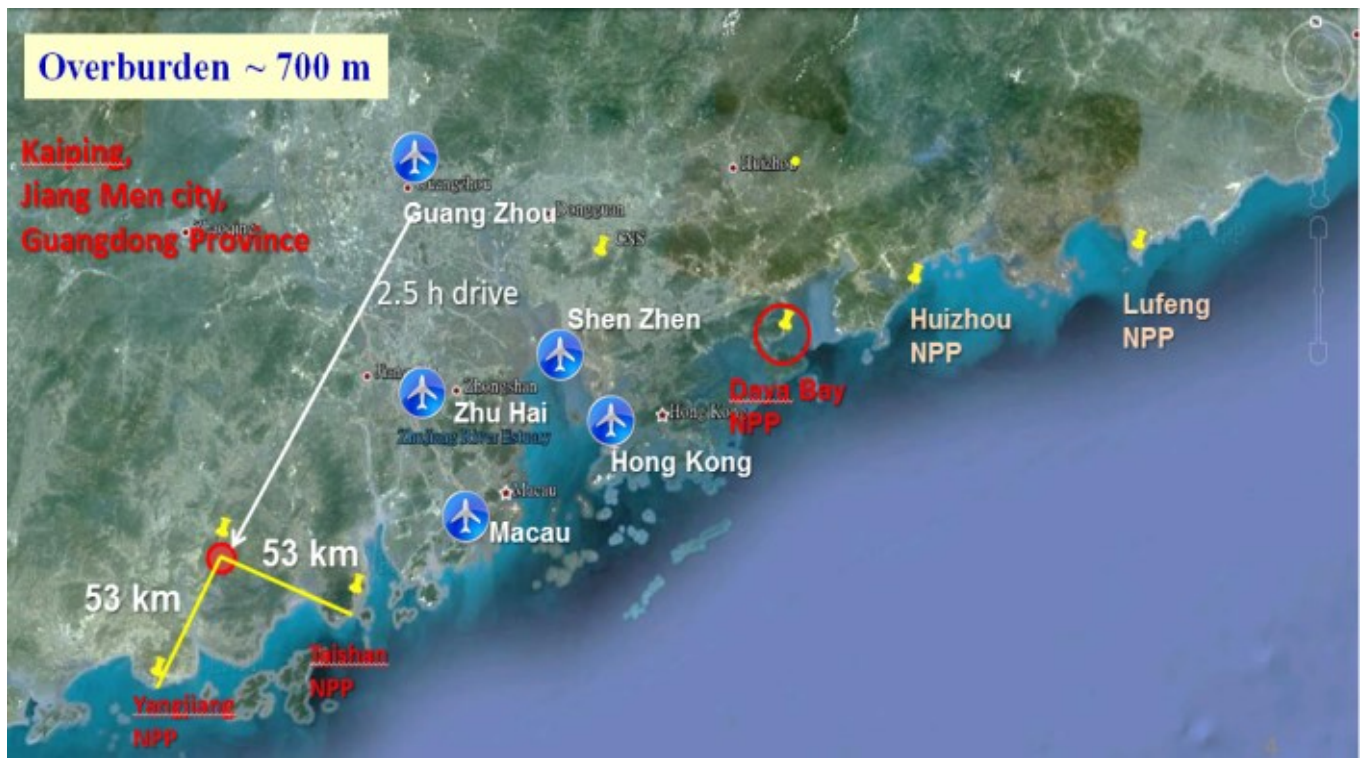

Fig.1. Default configuration of the JUNO experiment

Since the numerous PMTs with large-scaled front-end electronics readout channels in JUNO, the current data acquisition (DAQ) is designed to be composed of thousands of software processes to accomplish data-taking, online process and event storage. So, the scalability of the DAQ cluster is over several hundreds of computing nodes connected via high speed network to obtain powerful computing ability and satisfy the general requirement of high bandwidth.

## II. The Online Software

The Online Software, a distributed system, which is designed to control and monitor the whole DAQ system, plays important global roles during the whole data taking period in JUNO, including configuring, monitoring, controlling, multi-processes management, information sharing, in addition to, however, the access to the raw events. It is a customizable distributed framework, which provides essentially the 'glue' that holds the various sub-systems together [2] and makes them work coordinately. It provides interfaces to the Dataflow System (which is responsible for the transportation of the raw data from the readout drivers to mass storage) [3], to the remote web control system as well as the detector control system (DCS).

### A. *The Online software requirement*

#### 1) *Partition Requirement*

The Online Software is designed to be a common framework fit for different high energy physics experiments such as JUNO. So the notion of *Partition* is strongly advised in case the detectors are independent or the experiment requires multiple DAQ systems running concurrently. Partition is related to the organization and hierarchy of DAQ architecture.

This work was supported by Strategic Priority Research Program of the Chinese Academy of Sciences.
Jin Li is with the State Key Laboratory of Particle Detection and Electronics, Institute of High Energy Physics (IHEP), Chinese Academy of Sciences (CAS) and the University of Chinese Academy of Science (UCAS), Beijing 100049, China. (email: lijin@ihep.ac.cn)
Minhao Gu, Fei Li, Kejun Zhu are with the State Key Laboratory of Particle Detection and Electronics, Institute of High Energy Physics (IHEP), Chinese Academy of Sciences (CAS), Beijing 100049, China.

*2) High availability Requirement*

In a distributed system, the problem of single point of failure (SPOF), especially the SPOF of the core components has to be solved to improve the availability of the online software system and even the availability of DAQ.

*3) Functional requirement*

*a) Configuration of application's deployment*

*b) Run Control architecture*

*c) Information sharing and database*

### *B. SOA introduction*

The general definition of Service Oriented Architecture (SOA) is a deployment methodology relying on the integration and interaction between loosely coupled services. In simple terms, it is a software architecture that treats any running application as a service. This design concept is popular in recent years, for SOA makes the distributed system flexible, modular, and reusable. A Service is a well-encapsulated function unit that runs independently and can be accessed remotely.

### *C. Technology selection*

*1) Communication layer*

As SOA is service oriented, that all functional units act as service module and communication between services is by way of message exchange [4], the underlying communication layer should be considered first.

After comparing a number of communication libraries available on the open software market, ZeroMQ[5] is chosen as the implementation of the network communication layer [6] to construct the message model to transmit data among distributed services. ZeroMQ's application programming interface (API) seems like traditional Berkeley socket, encapsulates low-levels statement and error handling complexities, however. Several typical and advanced communication models are provided: request-reply, publish-subscribe and push-pull. Different services could use different communication models.

*2) Serialization tools*

Protocol Buffers[7], developed by Google, is a language-neutral, platform-neutral, extensible way of serializing structured data - think XML, but smaller, faster, and simpler for use in communications protocols, data storage, and more. It provides easy and multi-language supported serialization method to allow the users to define unique ZeroMQ messages to transport among services.

*3) High availability tools*

When it comes to high availability (HA), Master/Slave models is the primary solution to be considered. But how to monitor the failover and make the slave take over the task that matters. Zookeeper [8], a reliable, scalable distributed coordination system widely used in big data framework such as Hadoop, could solve the problems and act as the service broker in SOA based online software.

## III. ONLINE SOFTWARE ARCHITECTURE DESIGN

### *A. The Software Architecture*

Fig 2 shows the SOA based design of online software. The Green part consists of all online software services discussed in the paper, providing functionality for external components such as the orange parts-user apps or upper services. Table 1 gives an overview of the core services of the online software which will be described below in detail.

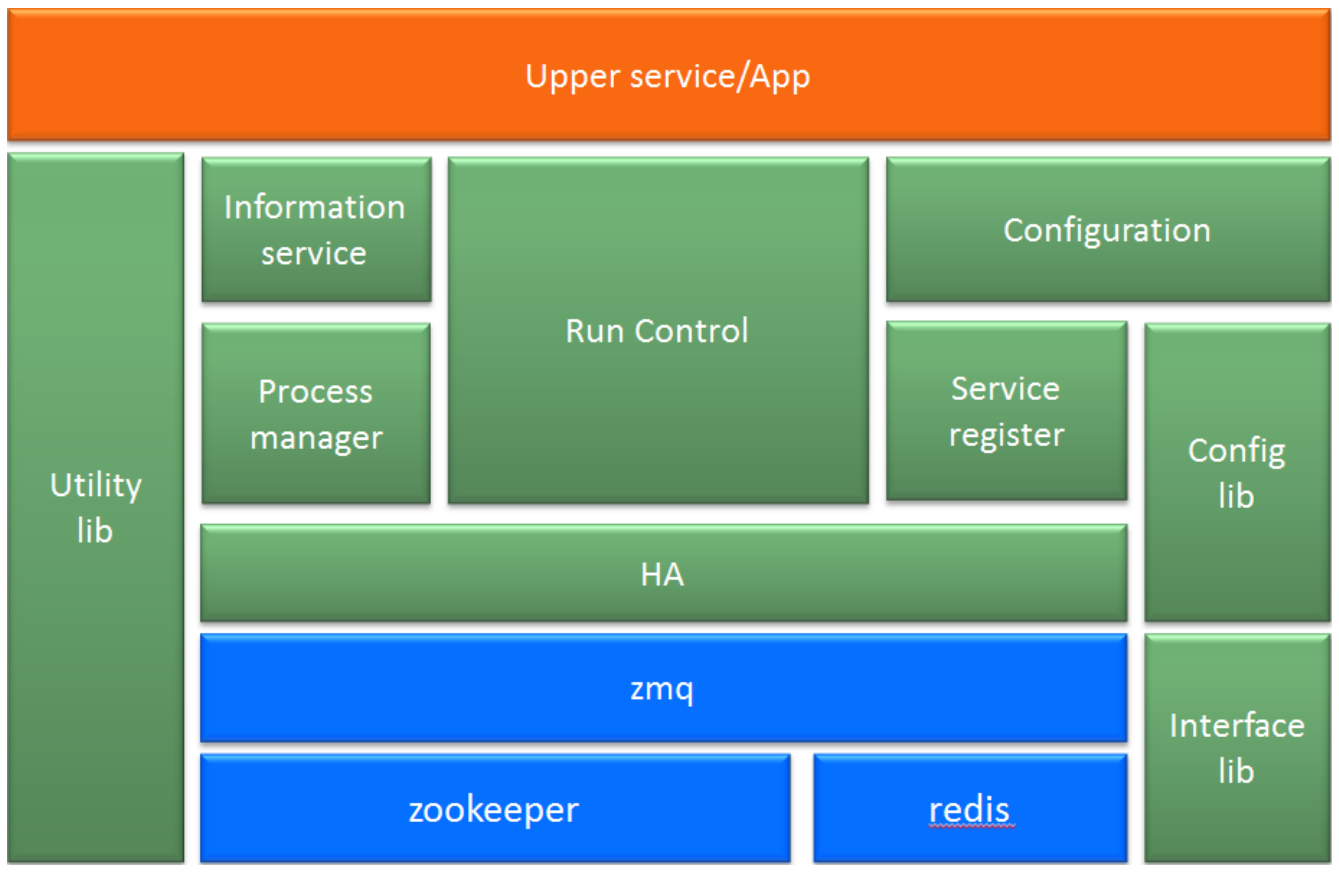

Fig.2. The SOA based design of online software

TABLE I
CORE SERVICES IN ONLINE SOFTWARE FRAMEWORK

| Service Name | Function description |
|---|---|
| *Configuration* | All software(services and applications) and hardware deployment |
| *Run Control* | Transaction of commands and states, with partition notation and hierarchical structure |
| *Information service* | Provides ways to store and access DAQ runs data, such as event rate and histograms |
| *Process Manager* | Runs on every node, provides management and monitor service for processes runs on the node |
| *High availability* | Keep every core services or applications accessible when error occurs |

### *B. Tree-like Run Control*

The Run Control service of the Online Software supplies all the necessary control and supervision for data taking by coordinating the different DAQ subsystem and detector operations from user interactions[9].The Run Control is the key service which in charge of all services and apps through issuing commands and obtaining states.

In order to satisfy the requirement of concurrent DAQs, the concept of 'partition' and 'segment' are used to divide the system into several sections. From Fig 3, several partitions are accessible at the same time while the "*Initial partition*" is a common partition responsible for initialization and public resource. The top-level RootController takes the role of overall control and coordination of the system. Each controller controls the direct next level's controllers, and fans recursively till the lowest level is apps or services.

Additionally, the Process Manager (PMG) is a service that using original process control provided by UNIX system. It includes the creation of new processes, program execution, and process termination.

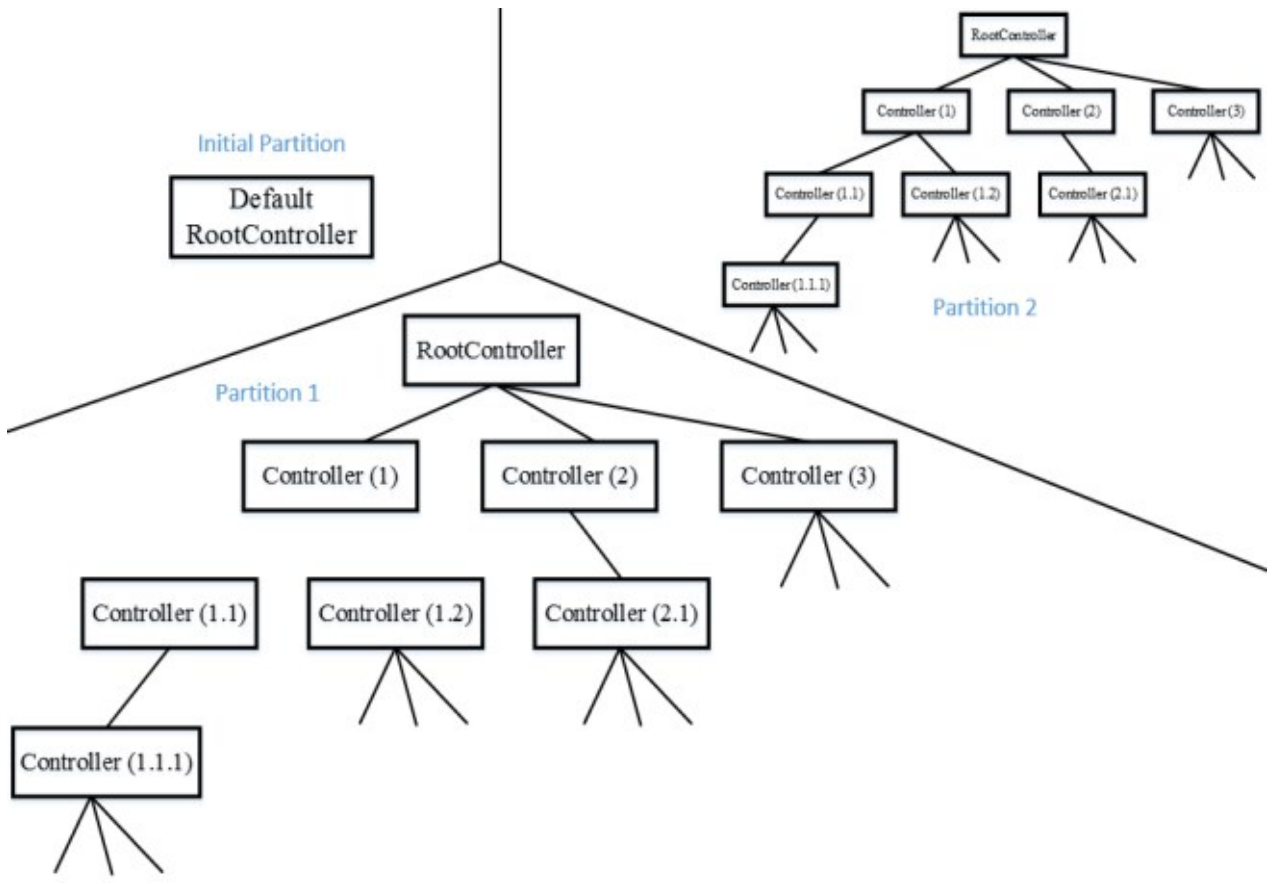

Fig.3. A hierarchical tree structure of controllers

### C. *Deployment Configuration*

Service of Configuration which is widely used by other services transfers user-defined hardware and software configuration parameters to internal memory. The user-defined configuration data are stored with XML format and CodeSynthesis XSD[10] is used for mapping XML file to C++ classes by predefined XSD schema file. The keyword '*IsControlledBy=*' that appears in the XML files means the object belongs to the management domain of the value which corresponding to Run Control service's *partition* concept and '*InfoDestTo*' means the information service destination which will be talked about later. This service also provides several request patterns according to id, type, regular-expression for the user.

The example XML codes are below:

```
<Object type="Partition" id="part_tst">
    <Rel name="Segment" type="segment" id="ros_seg" />
    <Rel name="Segment" type="segment" id="eb_seg" />
    <Rel name="Segment" type="segment" id="dfm_seg" />
    <Rel name="Segment" type="segment" id="setup" />
    <Rel name="IsControlledBy" type="service" id="RootController" />
    <Rel name="InfoDestTo" type="service" id="def_iss"/>
</Object>
```

### D. *High Availability*

In the DAQ system, some core components such as run control service should have the ability to auto-recovery from the failover. Aim at solving single-point failure in distributed services, Zookeeper is adopted. A simplest way to imply HA is Master/Slave model with which slave could take over the task in effective time when master down.

In one word, zookeeper is in charge of keeping the map of available services and offer auto-discovery mechanism. Because of the official client library that zookeeper provides is inconvenient to use, HA library has been developed to implement the automatic registration, recursively creation, reconnection for error toleration and recovery.

### E. *Information sharing*

Information Service (IS) stores all running information during the DAQ not only the data-taking periods, which includes DAQ states, event rates, histograms, error information and so on. Any service could publish its information to IS and any component could query information from IS. In order to take the pressure off the servers, the policy of load balance is designed.

In the underlying level, Redis has the advantages of fast, atomic, supporting multiple types of data to implement Information services. IS is also an interface service through which monitor service or DCS could get the DAQ's real-time running information. The Redis performance's test is shown in Fig.4. The result shows one Redis server could process at least 200K commands per second.

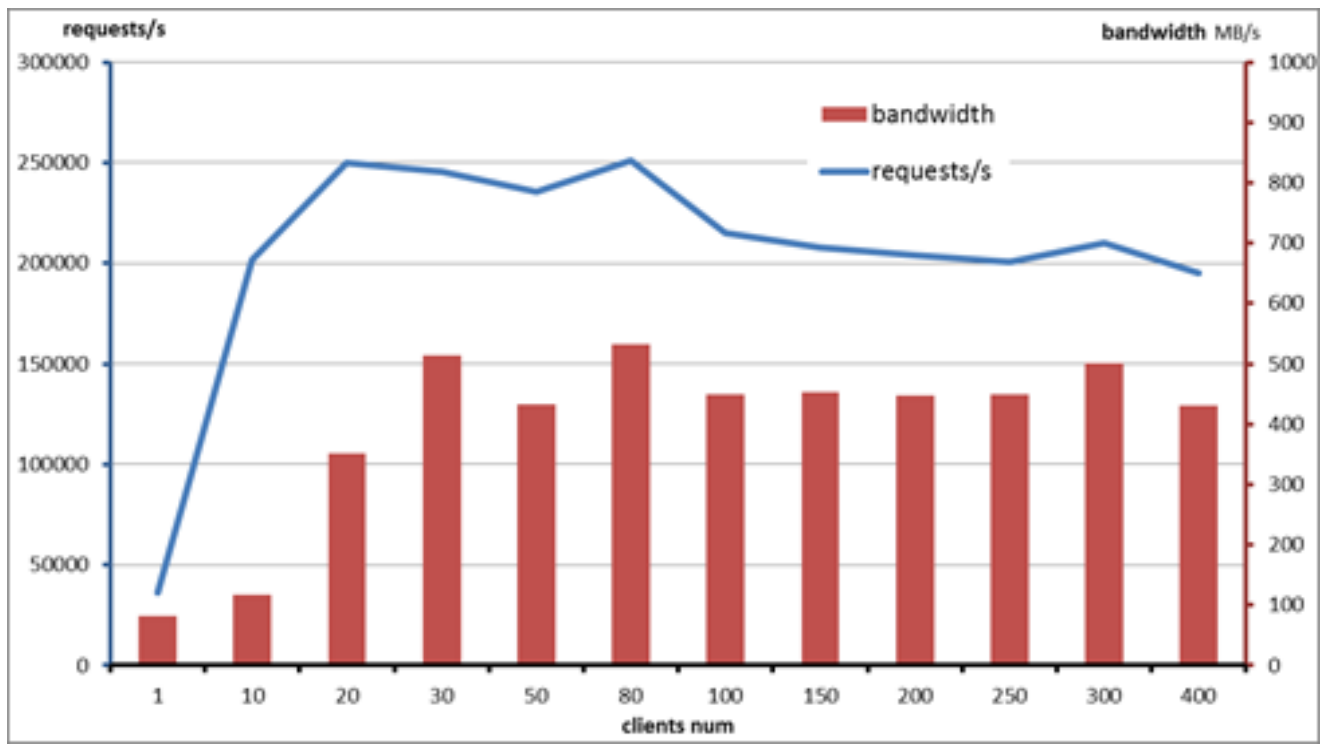

Fig.4. The blue line is the number of command which were successfully processed and returned per second for n clients. The red histograms are the Redis server's bandwidth for n clients.
The test environment is: the length of each request is 2000 Bytes. And 4 commands are wrapped in one request which is also called pipeline =4.

## IV. CONCLUSION

The architecture design of JUNO DAQ online software has been achieved after a comprehensive consideration of the modern IT technology and philosophy. The SOA and messages based design would be a new trend in high energy physics DAQ system.

However, the JUNO experiment onsite is under civil construction, so the DAQ software deployment is still in planning. The current step of the design is in the laboratory and the further development and optimization are in progress. More detail performance test is ongoing.